\documentclass{article}
\usepackage{amsfonts,amssymb, amsmath}
\usepackage[all]{xy}
\usepackage[english]{babel}

\def\nn{\nonumber }
\def\bq{ \begin{equation} }
\def\eq{ \end{equation} }
\def\ben{ \begin{eqnarray} }
\def\en{ \end{eqnarray} }

\def\g{{\gamma}}

\newtheorem{prop}{Proposition}

\newtheorem{re}{Remark}

\begin{document}


\title{Deformations of Poisson brackets and the Kowalevski top.}

\author{Yu.A. Grigorev, A. V. Tsiganov\\
\it\small
St.Petersburg State University, St.Petersburg, Russia\\
\it\small e--mail: yury.grigoryev@gmail.com, andrey.tsiganov@gmail.com}

\date{}
\maketitle

\begin{abstract}
Deformations of the known polynomial Poisson pencils associated with the Kowalevski top are studied. As a byproduct we find new variables of separation from the one of the Yehia systems and new bi-Hamiltonian description of the Sokolov system.
 \end{abstract}

\section{Introduction}
\setcounter{equation}{0}

 In Hamiltonian mechanics any function $H$ on the phase space $M$ generates vector field $X$ describing a dynamical system
 \bq\label{hvf}
 X=PdH\,.
 \eq
Here $dH$ is a differential of $H$, and $P$ is a Poisson bivector on $M$. This dynamical system is called integrable by Liouville if there exist a sufficient number of functionally independent functions $H_i$ on $M$ whose pairwise Poisson brackets are equal to zero
\bq\label{inv-gen}
\{H_i,H_k\}=(PdH_i,dH_k)=0\,.
\eq
In Hamiltonian dynamics, integrable systems are rather the exception than the rule. Still, within this celebrated class of Hamiltonian systems one encounters a whole hierarchy of possibilities. An important aspect is always how the dynamics behave under different perturbations.

For instance, in bi-Hamiltonian mechanics \cite{mag97} we are looking for second Poisson bivector $P'$ compatible with $P$ such that
 \bq\label{geom-eq1}
\{H_i,H_k\}'=(P'dH_i,dH_k)=0\,.
\eq
Starting with a given pencil of Poisson bivectors
\[
P_\lambda=P+\lambda P'\,,\qquad \lambda\in \mathbb C\,,
\]
we can study its possible Poisson deformations \cite{lih77}
\[P_\lambda\to \widetilde{P}_\lambda\]
and describe the corresponding perturbations of functions $H_i\to \widetilde{H}_i$, which have to be in involution with respect to deformed Poisson brackets.

The main aim of this paper is to study trivial de\-for\-ma\-tions  \cite{lih77} of two types
\ben
&I.\qquad &\widetilde{P}_\lambda =( P+\lambda P') +\lambda \mathcal L_{Y}P\,,\label{poi2-1}\\
\nn\\
&II. \qquad &\widetilde{P}_\lambda= \mathcal L_{Y}P+\lambda(P'+\Delta P'), \label{poi2-2}
\en
 associated with the Kowalevski top \cite{kow89}. Here $\mathcal L_Y$ is a Lie derivative along the vector field $Y$ such that $\mathcal L_Y{P}$ is still a Poisson bivector compatible with the second term at the pencil.

Recall that Kowalevski top is defined by the following Hamilton function
\bq\label{ham-kow-ini}
H_1=J_1^2+J_2^2+2J_3^2+2ax_1\,,\qquad a\in\mathbb R\,.
\eq
Here two vectors $J=(J_1,J_2,J_3)$ and $x=(x_1,x_2,x_3)$ are coordinates on the phase space $M$, which as a Poisson manifold is identified with the Euclidean algebra $e(3)^*$ with the Lie-Poisson brackets
\begin{equation}\label{e3}
\,\qquad \bigl\{J_i\,,J_j\,\bigr\}=\varepsilon_{ijk}J_k\,, \qquad
\bigl\{J_i\,,x_j\,\bigr\}=\varepsilon_{ijk}x_k \,, \qquad
\bigl\{x_i\,,x_j\,\bigr\}=0\,,
\end{equation}
where $\varepsilon_{ijk}$ is the totally skew-symmetric tensor.
These brackets have two Casimir functions
\begin{equation}\label{caz0}
C_1=\sum_{k=1}^3 x_k^2, \qquad
 C_2=\sum_{k=1}^3 x_kJ_k.
\end{equation}
Fixing their values one gets a generic symplectic leaf of $e^*(3)$ which is a four-dimensional symplectic manifold or a phase space for the dynamical systems under consideration \cite{bm05}. 

There is a large body of literature dedicated to the Kowalevski top, including the study of its integrable perturbations.
In \cite{bog84} Bogoyavlensky presented equations of motion for the top in two constant fields.
Komarov \cite{kom87} and, independently, Yehia \cite{yeh87} found the gyrostat extension of the Kowalevski top in 1987.
 In \cite{brs89} a Lax representation for the Kowalevski gyrostat in a double field was found by Bobenko, Reyman and Semenov-{T}ian-{S}hansky. In 2002, new integrable problems for the Kowalevski type gyrostat were found by Sokolov in the case of one axially symmetric field \cite{sok02} and by Sokolov and Tsiganov for the case of a double field \cite{sokts02}.

It is obvious that the Kowalevski top is integrable for the zero value of the square integral $C_2=0$, which is one of the Casimirs. In this case we have Kowalevski system on the cotangent bundle $T^*\mathbb S^2$ of the two-dimensional  sphere. First generalizations of the corresponding Hamiltonian were obtained by Chaplygin \cite{ch03} and Goryachev \cite{gor16}. An exhaustive list of all the known integrable perturbations of this particular system may be found in the Yehia papers, see for instance \cite{yeh06,yeh13}.

An abundance of many well-known integrable perturbation was a reason for the choice of the Kowalevski top for studying of the different Poisson pencils and deformations thereof. In fact, we have not a theory of such deformations till now. So, a collection of examples we present in this paper may be, in our opinion, regarded as a first step to such theory.

This paper is organized as follows. In the second part of Section 1 we will try to explain how deformations of the Poisson structures arise in an integrable systems theory in a natural way. This section is intended for an introduction to Section 2, which deals with the particular case of the Kowalevski top at $C_2=0$. In Section 2 we will give an explicit formula for the additive deformations (\ref{poi2-1}) pointing out some algorithm for their computation. Two known polynomial Poisson pencils $P_\lambda$ are the initial data for this algorithm and two new Poisson pencils $\tilde{P}_\lambda$ are its results.
As a byproduct we will obtain variables of separation from one of the Yehia system.

 Section 3 is devoted to the trivial deformations of the second type (\ref{poi2-2}). In this section we will give new deformations of the polynomial Poisson pencil associated with the so-called Kowalevski gyrostat with the double force field. As a byproduct we will obtain new rational Poisson structure in the Sokolov case. In conclusion, we discuss the construction of the bi-Hamiltonian structure for the Kowalevski top on $so^* (4) $.

\subsection{Additive deformations}
The explicit knowledge of first integrals $H_i$ of a given Hamiltonian system allows us to construct its infinitesimal symmetries
\[
X_i=PdH_i\,.
\]
Recall that by infinitesimal symmetry of a given dynamical system $\dot{x} = X$ we mean a smooth vector field $Y$ that
commutes with $X$, i.e., $[X, Y ] = 0$, or equivalently $\mathcal L_X Y = 0$.

Thus, according to \cite{mr}, one immediately gets a family of rank-two compatible Poisson structures for any integrable Hamiltonian system
\[
P^{(ij)}=X_i\wedge X_j\,.
\]
Recall that bivector $P$ is a Poisson structure if and only if
\bq\label{mY-eq}
[\![P,P]\!]=0
\eq
 Here $[\![.,.]\!]$ is a Schouten bracket defined by
\[
[\![P,P']\!]_{ijk}=-\sum\limits_{m=1}^{dim\, M}\left(P'_{mk}\dfrac{\partial P_{ij}}{\partial z_m}
+P_{mk} \dfrac{\partial P'_{ij}}{\partial z_m}+\mathrm{cycle}(i,j,k)\right)\,.
\]
In our case
\[
[\![P^{(i,j)},P^{(i,j)}]\!]=2X_i\wedge X_j\wedge [X_i,X_j]=0
\]
and integrals of motion $H_k$ are the Casimir functions of the corresponding Poisson bracket
\[
\{f,g\}^{(ij)}=\mathcal L_{X_i} f\cdot \mathcal L_{X_j}g-
\mathcal L_{X_i} g\cdot \mathcal L_{X_j}f\,.
\]
In generic every   pair of independent smooth vector fields $X_{1,2}$ such that
\bq\label{solv-x12}
 [X_1,X_2] = f_1X_1 + f_2X_2\,,\qquad f_{1,2}\in C^\infty(M),
 \eq
yields a Poisson structure because
\[
[\![P^{(1,2)},P^{(1,2)}]\!]=2X_1\wedge X_2\wedge [X_1,X_2]=2X_1\wedge X_2\wedge (f_1X_1+f_2X_2) =0\,.
\]
So, we have a lot of the Poisson pencils $P+\lambda P^{(i,j)}$, which could be natural initial data for the further deformations.

Namely, let us to suppose that  perturbed Hamilton  function is  a linear combination of the kinetic energy $T$ and potential energy $V$
\[\widetilde{H}=T+\mu V\,,\qquad \mu\in\mathbb R\,,\]
so that
\[\widetilde{X}_1=Pd\widetilde{H}=PdT+\mu PdV=X_T+\mu X_V\,. \]
If there is independent of $\mu$ vector field  $X_2$, so that equation (\ref{solv-x12}) holds  for initial and perturbed systems, then we have the following Poisson bivector
\bq\label{def-gen}
\widetilde{P}^{(12)}=\widetilde{X}_1\wedge X_2=X_T\wedge X_2+\mu X_V\wedge X_2= P^{(12)}_T+\mu P^{(12)}_V\,.
\eq
This bivector may be considered as deformation of the "kinetic" bivector $P^{(1k)}_T$ derived from two simple suggestions. It is easy to see that in this case coupling constant $\mu$ plays the role of a free parameter at the Poisson pencil.

So with that in mind, we can try to obtain perturbations of known integrable systems using  deformations of the Poisson bivectors. For instance, let us consider the free motion in the phase space $M=\mathbb R^{2n}$
\[
H=p_1^2+\cdots+p_n^2\,,\qquad X_1=PdH\,,\qquad
 P=\left(
 \begin{array}{cc}
 0\qquad&\mathrm I\\
 -\mathrm I&0
 \end{array}
 \right)
 \]
equipped with an additional vector field
 \[
 X_2=(q_1,\ldots, q_n,-p_1,\ldots,-p_n)
 \]
 such that
 \bq\label{comm-x12}
[X_1,X_2]=2X_1\,.
 \eq
 The corresponding  second rank-two Poisson structure has the following form
\bq\label{poi-2}
P'=X_1\wedge X_2=\mathcal L_{Z}\,P\,,\qquad  Z=(p_1q_1+\cdots+p_nq_n)PdH
\eq
This bivector has the Casimir function
\[
C=(p_1q_1+\cdots+p_nq_n)^2-(q_1^2+\cdots+q_n^2)H\,,\qquad P'dC=0\,,
\]
which satisfies to the separated relation
\[
r^2H+C=p_r^2r^2\,,
\]
where $r=\sqrt{q_1^2+\cdots +q_n^2}$ and $p_r$ is the conjugated momenta. It means that we consider a dynamical system partially separable in spherical coordinates.

According to \cite{ts13}  pencil $P+\lambda P'$ has a trivial deformation
\bq\label{y-gen}
\widetilde{P}_\lambda=P_\lambda+ \lambda\mathcal L_{Y} P\,,\qquad Y=(p_1q_1+\cdots+p_nq_n)PdV(q_1,\ldots,q_n)
\eq
where
\[
V(q_1,\ldots,q_n)=\sum_{i=1}^n \frac{1}{q_k^2}\,V_k\left(\frac{q_1}{q_k},\frac{q_2}{q_k},\ldots,\frac{q_n}{q_k}\right)\,.
\]
and $V_k$ are arbitrary functions on homogeneous coordinates $q_i/q_k$.  It is important that entries of the vector field $Z$ (\ref{poi-2}) are the second order polynomials in momenta, whereas entries of the vector field $Y$ (\ref{y-gen}) are the first order polynomials in momenta.

The corresponding perturbed integrals of motion
\bq\label{m-pot}
\widetilde{ H}=\sum_{i=1}^n p_i^2+V(q_1,\ldots,q_n)\,,\qquad \widetilde{C}=(p_1q_1+\cdots+p_nq_n)^2-(q_1^2+\cdots+q_n^2)\widetilde{H}
\eq
satisfy to the same separated relation
\[
r^2\widetilde{H}+\widetilde{C}=p_r^2r^2
\]
and
\[[\widetilde{X}_1,X_2]=2\widetilde{X}_1\,,\qquad \widetilde{X}_1=Pd\widetilde{ H}\]
similar to the initial vector field (\ref{comm-x12}). The Casimir function $\widetilde{C}$ coincides with the well-known Jacobi integral of motion, see \cite{alb98,bkm09}.

Other integrals of motion are the Casimir functions of the rank-two Poisson bivector
\bq\label{def-r}
\widetilde{P}'=P'+\mathcal L_{Y} P\,,
\eq
which can not be found in generic form for an arbitrary potential $V$.

If we postulate that our dynamical system is invariant with respect to translations, i.e. that
 Hamiltonian (\ref{m-pot}) is
 \bq\label{fin-H3}
\widetilde{H}=\sum_{i=1}^n p_i^2+\sum_{k=1}^{n-1}\frac{1}{(q_{k+1}-q_k)^2}\,U_k\left(\frac{q_2-q_1}{q_{k+1}-q_k},\ldots,\frac{q_n-q_{n-1}}{q_{k+1}-q_{k}}\right)
\eq
then there are additional linear in momenta integral of motion, additional symmetry and two  Poisson's bivectors
\[
\widetilde{H}_1=p_1+\cdots+p_n\,,\qquad \widetilde{X}_3=PdH_1\,,\qquad \widetilde{P}^{(32)}=\widetilde{X_3}\wedge X_2\,,\qquad \widetilde{P}^{(31)}=\widetilde{X_3}\wedge \widetilde{X}_1
\]
and a  solvable algebra of vector fields
\[
 [\widetilde{X}_1,X_2]=2\widetilde{X}_1\,,\qquad [\widetilde{X}_3,X_2]=\widetilde{X}_3\,,\qquad [\widetilde{X}_1,\widetilde{X}_3]=0\,.
\]
At $n=3$  there are enough vector fields and integrals of motion in order to prove integrability of the corresponding dynamical system using the Euler-Jacoby theorem or the Lie integrability theorem  \cite{koz13,ts13}.

In order to apply  Liouville's theorem we can calculate an algebra of integrals
\bq\label{alg-int}
\begin{array}{lll}
\{\widetilde{H}_1,\widetilde{H}\}=0\,,\qquad& \{\widetilde{H}_1,\widetilde{C}\}=\widetilde{H}_3\,,\qquad &\{\widetilde{H}_1,\widetilde{H}_3\}=2\widetilde{H}_1^2-6\widetilde{H}\,,
\\ \\
\{\widetilde{H},\widetilde{C}\}=0\,,\qquad &\{\widetilde{H},\widetilde{H}_3\}=0\,,\qquad &\{\widetilde{H}_3,\widetilde{C}\}=4\widetilde{H}_1\widetilde{C}\,,
\end{array}
\eq
and  find its central elements $\widetilde{H}$ and
\[\widetilde{H}_4=\widetilde{H}_3^2-4\widetilde{C}(\widetilde{H}_1^2-3\widetilde{H})\,,\]
which is a fourth order polynomial in momenta. It is easy to see that independent integrals of motion $\widetilde{H}_1,\widetilde{H}$ and $\widetilde{H}_4$ are in the bi-involution with respect to both Poisson brackets. An existence of independent additional Jacobi's integral $C$ allows us to say about superintegrable system with the Hamiltonian (\ref{fin-H3}).

 In similar manner at $n=3$ we can recover integrals of motion for the Calogero-Moser systems associated with classical root systems, integrals of motion for the Rosochatius system and Gaffet systems \cite{bkm09}. These integrals of motion are  the Casimir functions of $\widetilde{P}'$ (\ref{def-r}), which are the third, fourth and sixth order polynomials in momenta.

Summing up, we have proved that trivial deformation of the Poisson pencil could give us a new family of completely integrable systems. Of course,  instead of deformations  of rank-two Poisson bivectors $P^{(ij)}=X_i\wedge X_j $ we can directly study deformations of the corresponding vector fields $X_i$ and $X_j$. Therefore, in order to prove the efficiency of the theory of the Poisson deformations  below  we  will consider  rank-four and rank-eight Poisson vectors associated with the
 Kowalevski top and their trivial deformations.

\section{Polynomial Poisson bivectors}
\setcounter{equation}{0}
In this Section we consider Kowalevski top at the zero value of the square integral
\[C_2=(x,J)=0.\]
 At $C_2=0$ the corresponding symplectic leaf of $e^*(3)$ is symplectomorphic to $T^*{\mathbb S} ^2$ and, therefore, we can use standard spherical coordinate system on $T^*{\mathbb S}^2$ instead of variables $x$ and $J$
\bq\label{sph-coord}
\begin{array}{lll}
x_1 =\sin\phi\sin\theta,\qquad& x_2 = \cos\phi\sin\theta,\qquad & x_3 =\cos\theta\,,\\
\\
J_1 =\dfrac{\sin\phi\cos\theta}{\sin\theta}\,p_\phi-\cos\phi\,p_\theta\,,\qquad&
J_2 =\dfrac{\cos\phi\cos\theta}{\sin\theta}\,p_\phi+\sin\phi\,p_\theta\,,
\qquad& J_3 = -p_\phi\,.
\end{array}
\eq
 In these variables Poisson bivector
 \bq\label{can-p}
P=\left(
 \begin{array}{cc}
 0 & \mathrm I \\
 - \mathrm I & 0 \\
 \end{array}
 \right)
\eq
is nondegenerate and defines a symplectic form. Therefore,  below we will construct  different Poisson-Nijenhuis manifolds ($\omega N$ manifolds) \cite{fp02,mag97} starting with the common symplectic manifold $T^*\mathbb S^2$.

Based on the example from the previous Section we propose to study perturbations of a given integrable system using the following algorithm:
\begin{itemize}
  \item takes one of the known Poisson pencils for the initial integrable system with polynomials in momenta entries;
  \item  calculates its possible trivial deformations using special vector fields, whose entries are lower order polynomials in momenta;
  \item finds the corresponding perturbations of the initial integrals of motion, which have to be in the bi-involution with respect to deformed Poisson brackets.
\end{itemize}
For the Kowalevski top two rank-four Poisson structures $P'$ are known at $C_2\neq0$ and three rank-four Poisson structures at $C_2=0$. Four of them were obtained using Lax matrices for the Kowalevski top \cite{falq01,mar98,ts08} and one remaining was obtained directly from integrals of motion \cite{ts10k}. Entries of three bivectors $P'$ are rational functions in momenta, whereas entries of the two remaining bivectors are polynomials that allows us to apply  the proposed algorithm.

\subsection{First deformation}
In \cite{ts10k,ts11s} we found second Poisson bivector $P'$ for the Kowalevski top. Let us represent this bivector as a trivial deformation of $P$ similar to (\ref{poi-2})
\[
P'=\mathcal L_{Z}\,P\,,
\]
where
\[
Z=\left(
 \begin{array}{c}
 Z_T \\
 0 \\
 \end{array}
 \right)+\left(
 \begin{array}{c}
 0 \\
 Z_V \\
 \end{array}
 \right)=\left(
 \begin{array}{c}
 Z_T \\
 Z_V \\
 \end{array}
 \right)
\,.\]
Here a kinetic term is the second order polynomial in momenta
\bq\label{y-kow} Z_T=\left(
 \begin{array}{c}
 \left( \dfrac{1}{\cos\theta}+\ln\dfrac{1-\cos\theta}{\sin\theta}\right)p_\phi p_\theta \\ \\
 \dfrac{1}{2}\ln\dfrac{1-\cos\theta}{\sin\theta}\,p_\phi^2+\dfrac{1}{2\cos\theta} p_\theta^2 \\
 \end{array}
 \right)
\eq
and the potential term is the linear polynomial in momenta
\bq\label{y-kow-va}
Z_V= \dfrac{a}{2}\left(
 \begin{array}{c}
 \sin\phi\, p_\phi+\cos\phi\tan\theta\,p_\theta \\ \\
 -\cos\phi\cot\theta\,p_\phi -\sin\phi\,p_\theta\\
 \end{array}
 \right)\,.
\eq
 Now we can try to get deformation $\widetilde{P}_\lambda$ (\ref{poi2-1}) of the "kinetic" Poisson pencil $P_\lambda$
 \[
 P_\lambda=P+\lambda \mathcal L_Z P\,,\qquad Z=\left(
 \begin{array}{c}
 Z_T \\
 0 \\
 \end{array}
 \right)
 \]
 using the following ansatz for the entries of the vector field $Y$
 \bq\label{lin-anz}
 Y_i=f_i(\phi,\theta)p_\phi+g_i(\phi,\theta)p_\theta+h_i(\phi,\theta)\,.
\eq
Here $f_i, g_i$ and $h_i$ are some functions on the Euler angles. We use linear polynomials in momenta in order to get perturbations of potential in the initial Hamiltonian (\ref{ham-kow-ini}) only.

\begin{prop}
For the given kinetic part $Z_T$ (\ref{y-kow}) equation (\ref{mY-eq}) for $\widetilde{P}_\lambda$ has only two linear in momenta solutions
\[
Y=\left (
 \begin{array}{c}
 0 \\
 Y_V \\
 \end{array}
 \right)
\]
up to canonical transformations $\phi\to\phi+\alpha$
\[ Y_V^{(1)}=
\left(\begin{array}{c}
\frac{f'(\phi)}{\cos\theta}\\ \\
\frac{c_1-f(\phi)}{\sin\theta}
\end{array}\right) p_\theta
\]
and
\bq\label{first-dkow}
Y_V^{(2)}= c_1\left(
 \begin{array}{c}
 \sin\phi\, p_\phi+\cos\phi\tan\theta\,p_\theta \\ \\
 -\cos\phi\cot\theta\,p_\phi -\sin\phi\,p_\theta\\
 \end{array}
 \right)+\frac{c_2\,p_\theta}{\cos^2\phi}\left(
 \begin{array}{c}
 \frac{\tan\phi}{\cos\theta} \\ \\
 -\frac{1}{2\sin\theta} \\
 \end{array}
 \right)
 +\frac{c_3\,p_\theta}{\cos^2\phi}\left(
 \begin{array}{c}
 \frac{\cos^2\phi-2}{\cos\phi\cos\theta} \\ \\
 \frac{\sin\phi}{\sin\theta} \\
 \end{array}
 \right)
\eq
Here $f(\phi)$ is an arbitrary function and $c_k\in\mathbb R$ are arbitrary parameters.
\end{prop}
The proof is a direct solution of the overdetermined system of nonlinear differential equations up to canonical transformations.

In order to get integrals of motion $\widetilde{H}_{1,2}$ associated with $\widetilde{P}_\lambda$ we have to solve equations (\ref{inv-gen}-\ref{geom-eq1})
\bq\label{h12-eq}
(\widetilde{P}_\lambda\, d\widetilde{H}_1, d \widetilde{H}_2)=0\,,\qquad \forall \lambda,
\eq
with respect to $\widetilde{H}_1=H_1+\Delta H_1$ and $\widetilde{H}_2=H_2+\Delta H_2$, where perturbations $\Delta H_1$ and $\Delta H_2$ are first and third order polynomials in momenta, respectively.

First solution $Y_V^{(1)}$ of the equation (\ref{mY-eq}) yields integrals of motion
\[
\widetilde{H}_1=J_1^2+J_2^2+2J_3^2+v(x_3)\,,\qquad \widetilde{H}_2=J_3\,,
\]
whereas second solution at $c_1=a/2$, $c_2=-d$ and $c_3=e$ is associated with the Hamilton function
\[
\widetilde{H}_1=J_1^2+J_2^2+2J_3^2+2ax_1-\dfrac{2(x_2J_1-x_1J_2)\sqrt{b\,}}{x_3\sqrt{x_1^2+x_2^2}}+\dfrac{c}{\sqrt{x_1^2+x_2^2}}+\dfrac{2C_1-x_3^2}{x_2^2}\left(d+\dfrac{ex_1}{\sqrt{x_1^2+x_2^2}}\right)\,.
\]
Where $b$ is an additional parameter which arise in the solution of equation (\ref{h12-eq}). This Hamiltonian
after canonical transformation
\[
J_1=J_1-\dfrac{\sqrt{b\,}x_2}{x_3\sqrt{x_1^2+x_2^2}}\,,\qquad J_2=J_2+\dfrac{\sqrt{b\,}x_1}{x_3\sqrt{x_1^2+x_2^2}}\,,
\]
coincides with a well-known deformation of the Kowalevski top
\bq\label{ham-kow}
\hat{H}_1=J_1^2+J_2^2+2J_3^2+2ax_1-\dfrac{b C_1}{x_3^2}+\dfrac{c}{\sqrt{x_1^2+x_2^2}}+\dfrac{2C_1-x_3^2}{x_2^2}\left(d+\dfrac{ex_1}{\sqrt{x_1^2+x_2^2}}\right)
\eq
 First singular term $x_3^{-1}$ was added by Goryachev in \cite{gor16}. Other terms were added by Yehia, see \cite{yeh06}, recent paper \cite{yeh13} and references within.

This Hamiltonian (\ref{ham-kow}) commutes with
\ben \label{h2-kow}
\hat{H}_2&=&\left(J_1^2-J_2^2-2ax_1+\dfrac{b(x_1^2-x_2^2)}{x_3^2}\right)^2
+\left(2J_1J_2-2ax_2+\dfrac{2b x_1x_2}{x_3^2}\right)^2\\
\nn\\
&+&\dfrac{1}{x_2^4}\left(dx_3^2
+\dfrac{cx_2^2+ex_3^2x_1}{\sqrt{x_1^2+x_2^2}}\right)\left(2x_2^2(J_1^2+J_2^2)+dx_3^2
+\dfrac{cx_2^2+ex_3^2x_1}{\sqrt{x_1^2+x_2^2}}\right)\nn\\
\nn\\
&-&\dfrac{4ax_3^2(dx_1+e\sqrt{x_1^2+x_2^2})}{x_2^2}
-\dfrac{2b}{x_2^2}\left(\dfrac{\sqrt{x_1^2+x_2^2)}(cx_2^2-ex_3^2x_1)}{x_3^2}-dx_1^2\right)\nn
 \en
 with respect to the Poisson bracket (\ref{e3}).

So we do not get new integrable deformation of the Kowalevski system using this trivial deformation of the Poisson structures. Nevertheless, we find unknown bi-Hamiltonian structure of this system and, moreover, as a byproduct we can calculate unknown variables of separation for these known integrals of motion.

Recall that desired variables of separation are eigenvalues of the recursion operator \cite{fp02}
\[
N=(\widetilde{P}_\lambda-P)P^{-1}
\]
associated with deformation $\widetilde{P}_b$. In our case coordinates of separation $q_{1,2}$ are the roots of the characteristic polynomial \[\det(N-\mu I)=\det(P'P^{-1}-\mu \mathrm I)=B^2(\mu),\] where
 \ben
B(\mu)&=&(\mu-q_1)(\mu-q_2)\
=\mu^2+\left(\dfrac{\sqrt{x_1^2+x_2^2}(J_1^2+J_2^2)}{x_3^2}+\dfrac{ex_1+d\sqrt{x_1^2+x_2^2}}{x_2^2}\right)\mu\nn\\
\nn\\
&-&\dfrac{a(ax_3^2+2x_1J_1^2-2x_1J_2^2+4x_2J_1J_2)}{2x_3^2}-\dfrac{a(dx_1+e\sqrt{x_1^2+x_2^2})}{x_2^2}\,.
\label{q12-kow}
\en
They take values only in the following intervals
\[ q_1>a>q_2\,,\]
similar to the standard elliptic coordinates on the sphere \cite{ts11r}. Then we can introduce first order in $\mu$ polynomial \ben
 A(\mu)&=&\dfrac{(\mu-q_1)p_2(q_2^2-a^2)}{q_2-q_1}
 +\dfrac{(\mu-q_2)p_1(q_1^2-a^2)}{q_1-q_2}=-\dfrac{x_1J_2-x_2J_1}{x_3}\mu-\dfrac{a\sqrt{x_1^2+x_2^2}\,J_2}{x_3}\,,\nn\\
 \nn\\
 &=&-\mu \tan\theta p_\theta -a\cos\phi\, p_\phi-a\sin\phi\tan\theta\,p_\theta \label{p12-kow}
\en
such that for any $\nu$ and $\mu$ we have
\[
\{B(\nu),
A(\mu)\}=\dfrac{1}{\mu-\nu}\,\Bigl((\mu^2-a^2)B(\nu)-(\nu^2-a^2)B(\mu)\Bigr)\,,\qquad
\{A(\nu),A(\mu)\}=0\,.
\]
These two relations guarantee that variables
\bq\label{p-kow}
p_{j}=\dfrac{1}{q_j^2-a^2}\, A(\mu=q_j)\,,\qquad j=1,2,
\eq
are canonically conjugated momenta for the coordinates $q_{1,2}$, i.e. that
\[ \{q_i,p_j\}=\delta_{ij}\,,\quad \{q_1,q_2\}=\{p_1,p_2\}=0\,.\]
Initial variables in terms of $q_{1,2}$ and $p_{1,2}$ look like
\ben\label{zam-kow}
J_3&=&-\dfrac{x_1J_1+x_2J_2}{x_3}\,,\qquad x_3=\sqrt{1-x_1^2-x_2^2\,}\,,\\
\nn\\
J_1&=&\dfrac{(a^2-q_1^2)(a\sqrt{x_1^2+x_2^2}+x_1q_2)x_3p_1}{a\sqrt{x_1^2+x_2^2}(q_2-q_1)x_2}+
\dfrac{(q_2^2-a^2)(a\sqrt{x_1^2+x_2^2}+x_1q_1)x_3p_2}{a\sqrt{x_1^2+x_2^2}(q_2-q_1)x_2}\,,\nn
\nn\\
J_2&=&\dfrac{(a^2-q_1^2)q_2p_1+(q_2^2-a^2)q_1p_2)x_3}{a\sqrt{x_1^2+x_2^2}(q_2-q_1)}\,,\nn
\en
and
\ben
x_1&=&\dfrac{(a^2-q_1^2)(a^2-q_1q_2)p_1^2}{a(q_2-q_1)^2}-\dfrac{2(a^2-q_2^2)(a^2-q_1^2)p_1p_2}{a(q_2-q_1)^2}
+\dfrac{(a^2-q_2^2)(a^2-q_1q_2)p_2^2}{a(q_2-q_1)^2}\nn\\
\nn\\
&-&\dfrac{a\Bigl(d(q_1q_2+a^2)-ae(q_1+q_2)\Bigr)}{(a^2-q_1^2)(q_2^2-a^2)}\,,\nn\\
\nn\\
x_2&=&\dfrac{1}{a(q_2-q_1)^2}\sqrt{\dfrac{z_1z_2}{(a^2-q_1^2)(q_2^2-a^2)}}\,.\nn
\en
Here
\ben
z_1&=&((a-q_1)p_1+(q_2-a)p_2)^2(a+q_2)^2(a+q_1)^2+a^2(q_2-q_1)^2(d+e)\,,\nn\\
\nn\\
z_2&=&((a+q_1)p_1-(q_2+a)p_2)^2(a-q_2)^2(a-q_1)^2+a^2(q_2-q_1)^2(d-e)\,.\nn
\en
Integrals of motion $\widetilde{H}_{1,2}$ and variables of separation $(q_1,p_1)$ or $(q_2,p_2)$ lie on the two copies of the genus three algebraic curve defined by equation
\ben\label{seprel-kow}
&\Phi(q,p)=&\left(2(q^2-a^2)p^2+\widetilde{H}_1+\sqrt{\widetilde{H}_2}+2a\dfrac{da-eq}{q^2-a^2}\right)\times\qquad\\
\nn\\
&\qquad\qquad\,&\left(2(q^2-a^2)p^2+\widetilde{H}_1-\sqrt{\widetilde{H}_2}+2a\dfrac{da-eq}{q^2-a^2}\right)
-4q^2+4cq-8\sqrt{b}(q^2-a^2)p=0\,.\nn
\en
We remark that in this case methods of the Poisson geometry allow us to get not only integrable perturbations of the initial dynamical system, but also variables of separation for the perturbed system.

\subsection{Second deformation}

In \cite{ts02} we found other variables of separation for the Kowalevski top at $C_2=(x,J)=0$ and discovered second Poisson bivector $P'$ in \cite{ts08} using the reflection equation algebra for the corresponding Lax matrix.

In this paper, we represent known bivector $P'$ as trivial deformation (\ref{poi-2}) of $P$ with respect to special vector field $Z=Z_T+aZ_V$, polynomial in momenta. The "kinetic" part of this field is equal to
\[
Z_T=\left(
 \begin{array}{c}
 \dfrac{2p_\phi p_\theta \cos\theta}{\sin\theta} \\ \\
 2\mathrm i p_\phi p_\theta\\ \\
 \dfrac{p_\phi^3(\cos^2\theta-4)}{3\sin^2\theta} -p_\phi p_\theta^2\\ \\
 \dfrac{2\mathrm i p_\phi^3\cos\theta}{\sin^3\theta} +\dfrac{\cos^2\theta p_\phi^2p_\theta}{\sin^2\theta}-\dfrac{p_\theta^3}{3}
 \end{array}
 \right)\,.
\]
As above we start with the "kinetic" Poisson pencil
 \[
 P_\lambda=P+\lambda P'=P+\lambda \mathcal L_{Z_T} P
 \]
 and consider its deformation $\widetilde{P}_\lambda$ (\ref{poi2-1})
 using the following ansatz for the entries of the vector field $Y$
 \[
 Y_i=\sum_{k=0}^2 u_{ik}(\phi,\theta)p_\phi^kp_\theta^{2-k}+f_i(\phi,\theta)p_\phi+g_i(\phi,\theta)p_\theta+h_i(\phi,\theta)\,,
\]
We use the second order polynomials in momenta in order to get perturbations of potential in the initial Hamiltonian (\ref{ham-kow-ini}) only.

\begin{prop}
For the given kinetic part $Y_T$ (\ref{y-kow}) equation (\ref{mY-eq}) has only one quadratic in momenta solution
associated with the Kowalevski top
\ben
 Y&=&c_1\left(
 \begin{array}{c}
 0 \\ \\
 0 \\ \\
 2\mathrm i e^{-\mathrm i\phi}\sin\theta p_\phi \\ \\
 -2\mathrm i e^{-\mathrm i\phi}\cos\theta p_\phi
 \end{array}
 \right)+c_2\left(
 \begin{array}{c}
 -\frac{2\mathrm i e^{-\mathrm i\phi}(\cos\theta\sin\theta p_\theta+\mathrm i \cos^2\theta p_\phi-2\mathrm i p_\phi)}{\sin\theta} \\ \\
 0 \\
 -2\mathrm i \rho e^{-\mathrm i\phi}\sin\theta p_\phi \\ \\
 2\mathrm i \rho e^{-\mathrm i\phi}\cos\theta p_\phi
 \end{array}
 \right) \nn\\
\label{dy-cub}\\
 \nn \\
&+&c_3\left(
 \begin{array}{c}
 0 \\ \\
 0 \\ \\
 \frac{p_\phi-\rho}{\cos^2\theta} \\ \\
 \frac{\mathrm i(2\sin\theta p_\phi-\mathrm i \cos\theta p_\theta}{\cos^3\theta}
 \end{array}
 \right)-\rho\left(
 \begin{array}{c}
 \frac{2p_\theta\cos\theta}{\sin\theta} \\ \\
 0 \\ \\
 \frac{p_\phi^2(\cos^2\theta-2)}{\sin^2\theta} +\rho p_\phi-p_\theta^2\\ \\
 \frac{2p_\phi p_\theta \cos^2\theta}{\sin^2\theta}+\rho p_\theta
 \end{array}
 \right)\nn
\en
up to canonical transformations $\phi\to\phi+\alpha$.
\end{prop}
We have to underline that this solution nonlinearly depends on parameter $\rho$ in contrast with the previous cases (\ref{def-gen}) and (\ref{first-dkow}).

In this case deformation $\widetilde{P}_\lambda$ (\ref{poi2-1}) is associated with the Hamiltonian
\[
\widetilde{H}_1=J_1^2+J_2^2+2J_3^2+2\rho J_3-2c_1x_1+2\mathrm i c_2\Bigl(\rho x_2-(x_3J_2-2x_2J_3)\Bigr)-\dfrac{c_3}{x_3^2}\,,
\]
which after canonical transformation
\[
J_2 = J_2+\mathrm i c_2x_3\,,\qquad J_3 =J_3-\mathrm i c_2x_2\,,
\]
coincides with the Hamilton function for the Kowalevski-Chaplygin-Goryachev gyrostat \cite{kuzts89}
\[
\hat{H}=J_1^2+J_2^2+2J_3^2+2\rho J_3-2c_1x_1-c_2^2\left(x_1^2-x_2^2\right)-\dfrac{c_3}{x_3^2}
\]
At $\rho=c_2=c_3=0$ this system coincides with the Kowalevski case of motion of a rigid body about a fixed point \cite{kow89}. At $\rho=c_1=c_3=0$ it is the well-known case of Chaplygin in the dynamics of a rigid body moving by inertia in an infinitely extended ideal incompressible fluid \cite{ch03}. Potential term $x_3^{-1}$ was added by Gorychev in \cite{gor16}, whereas gyrostatic term $2\rho J_3$ was added by Komarov \cite{kom87} and by Yehia \cite{yeh87}.

As above, we do not get new integrable deformation of the Kowalevski system using trivial deformation of the Poisson structures. The main result is an explicit formula for the new bivector $\widetilde{P}'$, which is the 2-cocycle in the Poisson-Lichnerowicz cohomology defined by canonical Poisson bivector $P$ on $T^*\mathbb S^2$.

\section{Rational Poisson bivectors}
\setcounter{equation}{0}

For the Kowalevski top there are three Poisson bivectors $P'$ with rational entries in momenta. First of them associated with the Kowalevski variables of separation \cite{kow89} has been obtained using $2\times 2$ Lax matrix proposed in \cite{kuz02-lax} and  the corresponding reflection equation algebra \cite{ts08}. Second bivector $P'$ was obtained in \cite{mar98} and then in \cite{falq01} using $5\times5$ or $4\times 4$ Lax matrix constructed by Bobenko--Reyman--Semenov-Tian-Shansky \cite{brs89}. Third rational Poisson bivector was obtained using Lax matrices for the Kowalevski and Gorychev-Chaplygin gyrostats \cite{kuz02}.

Fortunately, one of them was obtained by the Dirac reduction procedure from the Lie-Poisson brackets (linear brackets) on extended 10-dimensional phase space, see details in \cite{falq01}. In this Section we discuss possible deformations of these linear brackets.

\subsection{Kowalevski gyrostat in a double field}
Our starting point is the final section of \cite{mar98}, where it has been shown that the Lax formulation \cite{brs89} for the so-called Kowalevski gyrostat in two fields with the Hamiltonian
\[
H=J_1^2+J_2^2+2J_3^2-2\rho J_3+2(x_1+y_2)
\]
admits both an $r$-matrix interpretation and a bi-Hamiltonian formulation. To achieve this, a peculiar splitting of the zero-degree part of the relevant twisted loop algebra, and an extension of the nine-dimensional Bobenko--Reyman--Semenov-Tian-Shansky phase space with variables $(x,y,J)$ to a ten-dimensional bi-Hamiltonian manifold with coordinates $(x,y,J,\varkappa)$ were performed. Recall that the Lax matrix
\[L=
\left(
 \begin{array}{ccccc}
 0 & J_3 & -J_2 &\lambda-\frac{x_1}{\lambda} &-\frac{y_1}{\lambda}\\ \\
 -J_3 & 0 & J_1 & -\frac{x_2}{\lambda} &\lambda-\frac{y_2}{\lambda}\\ \\
 J_2 &-J_1 & 0 &\frac{x_3}{\lambda}&-\frac{y_3}{\lambda} \\ \\
 \lambda- \frac{x_1}{\lambda} & -\frac{x_2}{\lambda} & \frac{x_3}{\lambda} & 0&\rho-J_3+\frac{\varkappa}{\lambda^2}\\ \\
 - \frac{y_1}{\lambda} &\lambda -\frac{y_2}{\lambda} & \frac{y_3}{\lambda} &J_3-\rho-\frac{\varkappa}{\lambda^2}& 0\\
 \end{array}
 \right)\,.
\]
satisfies the standard $r$-matrix algebra if the canonical Poisson brackets on ten-dimensional phase space are equal to
\bq\label{can-10}
\{J_i,J_j\}=\varepsilon_{ijk}J_k\,,\qquad \{J_i,x_j\}=\varepsilon_{ijk} x_k\,,\qquad
 \{J_i,y_j\}=\varepsilon_{ijk} y_k\,,\qquad
 \{x_i,y_j\}=\delta_{ij}\,\varkappa.
 \eq
The classical $r$-matrix in algebraic form is presented in \cite{brs89,mar98} and in matrix form in \cite{komts06}. Acting on this $r$-matrix $r_{12}(\lambda,\mu)$ by the simplest intertwining operator on the loop algebra, i.e. multiplying it on $\mu^{-2}$, one gets the second Poisson bivector
\bq\label{p2-ten}
P'=
 \left(\begin{smallmatrix}
0& J_3 &- J_2 &- J_3 +\rho&0&0&0&0&0& x_2 + y_1 \\
\noalign{\medskip}- J_3 &0& J_1 &0&- J_3 +\rho&0&0&0&1&- x_1 + y_2 \\
\noalign{\medskip} J_2 &- J_1 &0&0&0&- J_3 +\rho&0&-1&0& y_3 \\
\noalign{\medskip} J_3 -\rho&0&0&0& J_3 &- J_2 &0&0&-1&- x_1 + y_2 \\
\noalign{\medskip} 0& J_3 -\rho&0&- J_3 &0& J_1 &0&0&0&- x_2 - y_1 \\
\noalign{\medskip} 0&0& J_3 -\rho& J_2 &- J_1 &0&1&0&0&- x_3 \\
\noalign{\medskip} 0&0&0&0&0&-1&0&0&0& J_2 \\
\noalign{\medskip} 0&0&1&0&0&0&0&0&0&- J_1 \\
\noalign{\medskip} 0&-1&0&1&0&0&0&0&0&0\\
 - x_2 - y_1 & x_1 - y_2 &- y_3 & x_1 - y_2 & x_2 + y_1 & x_3 &- J_2 & J_1 &0&0
 \end{smallmatrix} \right) \,,
 \eq
which is independent on additional dynamic variable $\varkappa$.

We want to find all the possible linear deformations of the Lie-Poisson pencil $P+\lambda P'$. Therefore,  we have to substitute bivectors
\bq\label{tp-kow}
\widetilde{P}=P+\Delta P\,,\qquad \widetilde{P}'=P'+\Delta P'\,,
\eq
where $\Delta P$ and $\Delta P'$ are arbitrary linear functions on variables $x,y,J$ and $\varkappa$, into the Schouten brackets
\bq\label{meq-ten}
[\![\widetilde{P},\widetilde{P}]\!]=0\,,\qquad [\![\widetilde{P},\widetilde{P}']\!]=0\,,\qquad [\![\widetilde{P}',\widetilde{P}']\!]=0
\eq
and to solve the resulting equations with respect to $\Delta P$ and $\Delta P'$.

In our case   generic solution of the equations (\ref{meq-ten}) has the form (\ref{poi2-2}) and depends on two parameters $c$ and $\eta$
\[\widetilde{P}=\mathcal L_Y P\,,\]
where
\[\begin{array}{l}
Y_7=-J_1+c\sin\eta(\cos\eta x_3+\sin\eta y_3)\,,\qquad
Y_8=-J_2-c\cos\eta(\cos\eta x_3+\sin\eta y_3)\\ \\
Y_9=-J_3+c\cos\eta(\cos\eta x_2-\sin\eta x_1)+c\sin\eta(\cos\eta y_2-\sin\eta y_1)\,,\qquad
Y_{10}=\varkappa
\end{array}
\]
and other entries of $Y$ are equal to zero.

Below we put $\cos\eta=0$ for the brevity, so in this case second Poisson bivector looks like
\[
\widetilde{P}'=P'+c \left( \begin{smallmatrix} 0&- y_1 & x_3 &- x_2 &0&0&0&- J_2 &- J_3 &\varkappa\\
 \noalign{\medskip} y_1 &0& y_3 &0&- x_2 &0& J_2 &0&0&0\\
 \noalign{\medskip} - x_3 &- y_3 &0&- y_3 &0&- x_2 + y_1 &0&0& J_1 &0\\
 \noalign {\medskip} x_2 &0& y_3 &0&- y_1 &0& J_2 &0&0&0\\
 \noalign{\medskip} 0& x_2 &0& y_1 &0& y_3 &- J_1 &0&- J_3 &-\varkappa\\
 \noalign{\medskip} 0&0& x_2 - y_1 &0&- y_3 &0&0&0& J_2 &0\\
 \noalign{\medskip} 0&- J_2 &0&- J_2 & J_1 &0&0&0&0&0\\
 \noalign{\medskip} J_2 &0&0&0&0&0&0&0&0&0\\
 \noalign{\medskip} J_3 &0&- J_1 &0& J_3 &- J_2 &0&0&0&0\\
 \noalign{\medskip} -\varkappa&0&0&0&\varkappa&0&0&0&0&0
 \end{smallmatrix} \right)\,.
\]
This Poisson bivector can not be represented as the Lie derivative of $P$ or $P'$ along the vector field. It means that $\widetilde{P}'$  is nontrivial deformation \cite{lih77}.

The Hamilton function associated with this deformation of the Poisson pencil is a Casimir function of the second Poisson bivector $\widetilde{P}'$
\bq\label{tssok-ham}
\widetilde{H}=H+2c(J_1y_3-J_2x_3+J_3x_2-J_3y_1)\,.
\eq
This Hamiltonian was found in \cite{sokts02} together with the Lax matrices. The corresponding classical $r$-matrix is discussed in \cite{sok04,ts04} and the phase topology in \cite{ryab13}.

Here we do not discuss deformations of the Lax matrices and $r$-matrices on the extended phase space related to the obtained deformations of the Poisson brackets. We only present the two corresponding Lenard chains similar to \cite{falq01}:
\[
 \xymatrix{
 0 & & & \\
 d\widetilde{H_0} \ar[u]_{\widetilde{P}'} \ar[r]_{\widetilde{P}} & X_1& & \\
 &d\widetilde{H_1} \ar[u]_{\widetilde{P}'} \ar[r]_{\widetilde{P}}&X_2& \\
 & &d\widetilde{H_2}\ar[u]_{\widetilde{P}'} \ar[r]_{\widetilde{P}}&0
 }\qquad
 \xymatrix{
 0 & & & \\
 d\widetilde{K_0} \ar[u]_{\widetilde{P}'} \ar[r]_{\widetilde{P}} & Y_1& & \\
 &d\widetilde{K_1} \ar[u]_{\widetilde{P}'} \ar[r]_{\widetilde{P}}&Y_2& \\
 & &d\widetilde{K_2}\ar[u]_{\widetilde{P}'} \ar[r]_{\widetilde{P}}&0
 }\,.
\]
This diagram means that in the explicit form first Lenard chain reads as
\[
\widetilde{P}'d\widetilde{H}_0=0\,,\qquad X_1=\widetilde{P}'d\widetilde{H_1}=\widetilde{P}d\widetilde{H}_0\,,\qquad X_2=\widetilde{P}'d\widetilde{H_2}=\widetilde{P}d\widetilde{H}_1\,,\qquad \widetilde{P}d\widetilde{H_2}=0
\]
and similar to the second chain. First chain relates differentials of the following integrals of motion
\[
\widetilde{H}_0=\widetilde{H},\qquad\widetilde{H}_1=-(x,x)-(y,y)-2\varkappa(J_3-\rho+cx_2)\,,\qquad \widetilde{H}_2=\varkappa^2\,.
\]
Integrals of motion in the second Lenard chain are more lengthy
\ben
\widetilde{K}_0&=&2\Bigl(c^3J_2^2(y_1-x_2)-c^2J_2\bigl(J_2(J_3-\rho)-2y_3\bigr)+c(J_1J_2-x_2-y_1)+\rho\Bigr)\varkappa\nn\\
&+&(x_2-y_1)\Bigl (( x_2+y_1)(J_1^2-J_2^2)+( x_2- y_1)J_3^2 -2 J_1 J_2(x_1- y_2)+2 J_3 (J_1y_3-J_2 x_3)\Bigr) c^2\nn\\
&+&J_3^4+2(c x_2- c y_1- \rho ) J_3^3+\Bigl(J_3^2+2( c x_2- \rho ) J_3-2 c \rho x_2+2 x_1\Bigr) J_1^2\nn\\
&+&(J_3^2+(-2 c y_1-2 \rho ) J_3+2 c \rho y_1+2 y_2) J_2^2+\Bigl(2 c \rho(y_1- x_2)- \rho ^2+2 x_1+2 y_2\Bigr) J_3^2\nn\\
&+&2\Bigl(
\bigl(cJ_3(y_2-x_1)+c\rho(x_1-y_2)+x_2+y_1\bigr)J_2+(\rho-cx_2-cy_1)x_3+cy_3(J_3^2-\rho J_3-\rho^2+2x_1)
\Bigr) J_1\nn\\
&+&2\Bigl(y_3(cx_2+cy_1+\rho) -c(J_3^2-\rho J_3-\rho^2+2y_2)x_3\Bigr) J_2-2 ( \rho ^2-x_1-y_2) (c x_2-c y_1- \rho ) J_3\nn\\
&-&2 \rho ^2(x_1+y_2)+4 x_1 y_2-x_2^2-2 x_2 y_1-y_1^2\,,\nn
\en
\ben
\widetilde{K}_1&=&(1+c^2J_2^2)\varkappa^2-2\varkappa
\Bigl(J_3^3+ \rho ^3+(cx_2+J_3- \rho )J_1^2+(cx_2-cy_1+J_3- \rho )J_2^2+(cx_2-2cy_1- \rho )J_3^2\Bigr.\nn\\
&+&(c^2x_2y_3+J_2cy_2+2J_3cy_3-c \rho y_3-x_3)J_1
+(c^2x_1y_3-c^2x_3y_1+c^2y_2y_3-y_3)J_2\nn\\
&+&\Bigl.(c^2y_1^2+c^2y_3^2-c^2x_2y_1+c \rho y_1- \rho ^2+x_1+y_2)J_3
-c \rho ^2 x_2+cx_2y_2-cy_1y_2
\Bigr)+2\varkappa c(x,y)\nn\\
&-&(J_3-cy_1)(x\times y,J)-c\Bigl (x_1(x_1 y_3- x_3 y_1)+x_2(x_2 y_3- x_3 y_2)+y_1(x_2 y_3-x_3 y_2)\Bigr) J_1\nn\\
&-&c y_2 (x_2 y_3-x_3 y_2) J_2-c\Bigl (x_1 (x_2 y_2+ x_3 y_3)-y_1(x_2^2+x_3^2)+y_3(x_2 y_3-x_3 y_2)\Bigr) J_3\nn\\
&-&x_1^2 y_2+x_1 x_2 y_1-x_1 y_2^2-x_1 y_3^2+x_2 x_3 y_3+x_2 y_1 y_2-x_3^2 y_2+x_3 y_1 y_3+\rho^2\Bigl((x,x)+(y,y)\Bigr)\nn\\
&-&(x,J)^2-(y,J)^2-2\rho(x\times y,J)\,,\nn
\en

\ben
\widetilde{K}_2&=& \Bigl( c^2( y_1 ^2+ y_3 ^2)+2c( J_1 y_3- J_3 y_1)+ J_1 ^2+ J_2 ^2
+ J_3 ^2-\rho^2 \Bigr) \varkappa^2
+2 \Bigl( c y_1(x_2 y_1- x_1 y_2 )\Bigr. \nn\\
&+&\Bigl. cy_3( x_2 y_3 - x_3 y_2)+ J_1( x_2 y_3- x_3 y_2)+ J_2(
x_3 y_1- x_1 y_3)+ J_3( x_1 y_2- x_2 y_1) \Bigr) \varkappa
\nn\\
&+& (x,x)+(y,y)-(x,y)^2.\nn
\en

So, deformation of the Poisson pencil gives us known integrable system. The new result is a bi-Hamiltonian structure for this known system. Moreover, this example of the new type of deformations (\ref{poi2-2}) may be very useful for studying of the St\"ackel systems, which are bi-Hamiltonian systems namely on the extended phase space \cite{imm00}.

\subsection{Sokolov system}
In order to get the Sokolov system we have to make the canonical transformation
\[x_i\to ax_i
\,,\qquad y_i\to ay_i\,,\qquad \varkappa\to a^2\varkappa\]
and rescaling $c\to ca^{-1}$. Then we have to impose restrictions $y_1=0, y_2=0, y_3=0$ and $\varkappa=0$ on the phase space of the Kowalevski gyrostat (\ref{tssok-ham}) in two fields, that allows us to get the following Hamiltonian on $e^* (3) $
\bq\label{ham-sg}
\widetilde{H}=J_1^2+J_2^2+2J_3^2-2\rho J_3+2ax_1+2c(J_3x_2-J_2x_3)\,.
\eq
 At $a=0$ this Hamiltonian coincides with the Hamiltonian for the so-called Sokolov system \cite{bm05,sok02}
\[
\widehat{H}=J_1^2+J_2^2+2J_3^2-2\rho J_3+2c(J_3x_2-J_2x_3)\qquad \mbox{at}\quad a=0\,.
\]
It is easy to prove that applying canonical transformation and the Dirac procedure to the first bivector $\widetilde{P}$ (\ref{tp-kow}) one gets canonical Poisson brackets on $e^*(3)$ (\ref{e3}), see \cite{falq01} for the details. The second Poisson bivector $\widetilde{P}'$ (\ref{tp-kow}) becomes a linear combination
\bq\label{p2-gsok}
\widetilde{P}'=\widetilde{P}'_1+(x,J)^{-1}\widetilde{P}'_2
\eq
where
\[
\widetilde{P}'_1=\left( \begin {array}{cccccc}
0&{J_3}&c{x_3}-{J_2}&0&-c{J_2}&-c{J_3}\\
\noalign{\medskip}-{J_3}&0&{J_1}&c{J_2}&0&a\\
\noalign{\medskip}-c{ x_3}+{ J_2}&-{J_1}&0&0&-a&c{J_1}\,\\
\noalign{\medskip}0&-c{J_2}&0&0&0&0\\
\noalign{\medskip}c{J_2}&0&a&0&0&0\\
\noalign{\medskip}c{J_3}&-a&-c{J_1}&0&0&0
 \end {array} \right)
\]
and entries of the second matrix read as
\ben
(\widetilde{P}'_2)_{12}&=& \left( cx_2+J_3-\rho \right) \left( -cx_2 x_3+J_1 x_1+J_2 x_2-J_3 x_3+\rho x_3 \right)\nn\\
(\widetilde{P}'_2)_{13}&=& \left( cx_2+J_3-\rho \right) x_2 \left( cx_2+2 J_3-\rho \right) \nn\\ (\widetilde{P}'_2)_{14}&=&ax_2(cx_2+2J_3-\rho)-J_1(cx_2+J_3-\rho)(J_2-cx_3)\nn\\
 (\widetilde{P}'_2)_{15}&=&\left( cx_2+J_3-\rho \right) J_1^2\nn\\
 (\widetilde{P}'_2)_{16}&=&-ax_2J_1+c(cx_2+J_3-\rho)(x_2J_2+x_3J_3)\nn\\
 (\widetilde{P}'_2)_{23}&=&- \left( cx_2+J_3-\rho \right) x_1 \left( cx_2+2 J_3-\rho \right)\nn\\
 (\widetilde{P}'_2)_{24}&=&-ax_1(cx_2+2J_3-\rho)-J_2(cx_2+J_3-\rho)(J_2-cx_3)\nn\\
 (\widetilde{P}'_2)_{25}&=&\left( cx_2+J_3-\rho \right) J_1 J_2\nn\\
 (\widetilde{P}'_2)_{26}&=&a\bigl(x_1J_1-(cx_2+J_3-\rho)x_3\bigr)-cx_1J_2(cx_2+J_3-\rho)\nn\\
 (\widetilde{P}'_2)_{34}&=&
 - \left( cx_2+ J_3-\rho \right) \left( J_1 cx_1+J_2 cx_2+J_2 J_3 \right) \nn\\
 (\widetilde{P}'_2)_{35}&=& \left( cx_2+J_3-\rho \right) J_3 J_1\nn\\
 (\widetilde{P}'_2)_{36}&=&ax_2(cx_2+J_3-\rho)-cx_1J_3(cx_2+J_3-\rho)\nn \\
 (\widetilde{P}'_2)_{45}&=&aJ_3 J_1\nn\\
 (\widetilde{P}'_2)_{46}&=&a^2x_2+a\bigl(c(x_3J_1-x_1J_3)-J_1J_2\bigr)+c^2J_2C_2\nn\\
 (\widetilde{P}'_2)_{56}&=&aJ_1^2\nn
\en
  At $a=0$ Poisson bivector $\widetilde{P}'$ (\ref{p2-gsok}) defines  new rational bi-Hamiltonian structure for the Sokolov system. Recall that up to this time for the Sokolov system we have a polynomial Poisson pencil only at $C_2=0$ \cite{ts11r}.

At $a\neq 0$ we can use bivector $\widetilde{P}'$  (\ref{p2-gsok})  for the construction of unknown bi-Hamiltonian structure for the Kowalevski gyrostat on $so^*(4)$. This system is  defined by the Hamilton function
\[
\widetilde{H}_\kappa=J_1^2+J_2^2+2J_3^2-2\rho J_3+2x^{(\kappa)}_1\,,
\]
and the Poisson brackets
\begin{equation}\label{so4}
\,\qquad \bigl\{J_i\,,J_j\,\bigr\}=\varepsilon_{ijk}J_k\,, \qquad
\bigl\{J_i\,,x^{(\kappa)}_j\,\bigr\}=\varepsilon_{ijk}x_k^{(\kappa)} \,, \qquad
\bigl\{x_i^{(\kappa)}\,,x^{(\kappa)}_j\,\bigr\}=\kappa^2 \varepsilon_{ijk}J_k \,,
\end{equation}
In order to get this Hamiltonian and the desired bi-Hamiltonian structure we have to apply the Poisson map
\[J\to J\,,\qquad x\to x^{(\kappa)}=\alpha x+\gamma x\times J\,,\]
where $\alpha$ and $\g$ are special functions \cite{sokkom03}, to the Hamiltonian (\ref{ham-sg}) and to the  Poisson bivector $\widetilde{P}'$ (\ref{p2-gsok}).

The final expression for the second Poisson bivector on $so^*(4)$ is very bulky and, therefore, we don't provide it here. This rational bi-Hamiltonian structure and its possible deformations we will discuss in the separate publication.

\section{Conclusion}
 All the examples of integrable perturbations of the Kowalevski top we have discussed in this paper have appeared in the literature. We limit ourselves to give a new algorithm for construction of such perturbations and to present new examples of the compatible Poisson structures on $T^*\mathbb S^2$ and $e^*(3)$. We hope that acquired experience will be useful for studying deformations of the Poisson structures and new integrable perturbations of   integrable systems both on $T^*\mathbb S^2$ and $e^* (3) $.

We are grateful to the referees for a number of helpful suggestions for improvement in the article.
This work was partially supported by RFBR grant 13-01-00061.

\end{document}